\documentclass[twocolumn,showpacs,preprintnumbers,amsmath,amssymb,prb,superscriptaddress]{revtex4}

\usepackage{graphicx}
\usepackage{dcolumn}
\usepackage{bm}
\usepackage{upgreek}

\begin{document}


\title{Investigation of the chemical vicinity of crystal defects in ion-irradiated Mg and AZ31 with coincident Doppler broadening spectroscopy}


\author{M.~Stadlbauer}
\email[]{mstadlba@ph.tum.de}
\homepage[]{http://e21.frm2.tum.de/index.php?id=nepomuc}
\affiliation{Technische Universit{\"a}t M{\"u}nchen, Physikdepartment E21, James-Franck-Str., D-85747 Garching, Germany}
\affiliation{Technische Universit{\"a}t M{\"u}nchen, ZWE FRM II, Lichtenbergstr. 1, D-85747 Garching, Germany}

\author{C.~Hugenschmidt}
\affiliation{Technische Universit{\"a}t M{\"u}nchen, Physikdepartment E21, James-Franck-Str., D-85747 Garching, Germany}
\affiliation{Technische Universit{\"a}t M{\"u}nchen, ZWE FRM II, Lichtenbergstr. 1, D-85747 Garching, Germany}

\author{K.~Schreckenbach}
\affiliation{Technische Universit{\"a}t M{\"u}nchen, Physikdepartment E21, James-Franck-Str., D-85747 Garching, Germany}
\affiliation{Technische Universit{\"a}t M{\"u}nchen, ZWE FRM II, Lichtenbergstr. 1, D-85747 Garching, Germany}

\author{P.~B{\"o}ni}
\affiliation{Technische Universit{\"a}t M{\"u}nchen, Physikdepartment E21, James-Franck-Str., D-85747 Garching, Germany}

\date{\today}

\begin{abstract}
Crystal defects in magnesium and magnesium based alloys like AZ31 are of major importance for the understanding of their macroscopic properties. We have investigated defects and their chemical surrounding in Mg and AZ31 on an atomic scale with Doppler broadening spectroscopy of the positron annihilation radiation. In these Doppler spectra the chemical information and the defect contribution have to be thoroughly separated. For this reason samples of annealed Mg were irradiated with Mg-ions in order to create exclusively defects. In addition Al- and Zn-ion irradiation on Mg-samples was performed in order to create samples with defects and impurity atoms. The ion irradiated area on the samples was investigated with laterally and depth resolved positron Doppler broadening spectroscopy (DBS) and compared with preceding SRIM-simulations of the vacancy distribution, which are in excellent agreement. The investigation of the chemical vicinity of crystal defects in AZ31 was performed with coincident Doppler broadening spectroscopy (CDBS) by comparing Mg-ion irradiated AZ31 with Mg-ion irradiated Mg. No formation of solute-vacancy complexes was found due to the ion irradiation, despite the high defect mobility.

\end{abstract}


\maketitle

\section{Introduction}\label{Introduction}
Positron spectroscopy is a well-established tool to investigate lattice defects in solids.\cite{Dupasquier1995,West1973,PuskaNieminen1994, LynnSchulz1988} It has been widely used to investigate defects in metals \cite{Jackman1974} and precipitation processes in alloys like for example in the alloy Mg-Ca.\cite{Ortega2005} Coincident Doppler Broadening Spectroscopy (CDBS) with positrons gains more and more interest since it has proved to be a very sensitive technique to distinguish between different elements in the vicinity of crystal defects.\cite{Alatalo1995,AsokaKumar1996,Kuriplach1998} Thus, the formation process of embedded clusters in alloys like nuclear reactor pressure vessel steels \cite{Nagai2001,Nagai2004} or the binary alloy Al-Sn \cite{Cizek2005} were successfully investigated with CDBS.\\
After implantation into a sample, positrons thermalize within picoseconds and diffuse over hundreds of lattice spacings \cite{LynnSchulz1988} until they annihilate either directly with an electron or after being trapped in crystal defects. In the center of mass system, two $511\,\mathrm{keV}$ $\gamma$-quanta are emitted in opposite directions, whereas in the lab system the electron momentum causes a deviation from the $180^{\circ}$ angular correlation and an energy shift of the $511\,\mathrm{keV}$-quanta due to the Doppler effect (the momentum of the thermalized positron is negligible).\\
The shape of the $511\,\mathrm{keV}$-annihilation line reveals the electron momentum distribution at the annihilation sites due to the Doppler effect. Electrons with $7.08\,\mathrm{eV}$ \cite{Ashcroft1976} at the Fermi-level in magnesium for example cause a Doppler shift of $1.3\,\mathrm{keV}$ of the emitted annihilation quanta whereas e.g.\ the localized core electrons in the 2s shell with an binding energy of $88.6\,\mathrm{eV}$ \cite{Lide2003} lead to a Doppler shift of $4.8\,\mathrm{keV}$. Therefore, regions of the annihilation line with high Doppler shifts represent the momentum distribution of localized core electrons and hence allow a distinction between different elements. Since positrons are efficiently trapped in crystal defects,\cite{LynnSchulz1988} their chemical vicinity can be therefore measured with high accuracy. In the experiment the count rate in the high momentum region is low due to two facts: Firstly, in Doppler broadening (DB) or CDB measurements only the longitudinal projection of the electron momentum $p_l$ in one dimension is determined by measuring the energy shift $\Delta E = \frac{cp_l}{2}$ of the emitted photons.\cite{KrauseRehberg1999} Secondly, the core annihilation probability is in the range of typically a few \%.\cite{Jensen1990} Due to the low count rate in the high momentum regions, two facing high purity germanium detectors in coincidence instead of one have to be used in order to suppress the background by many orders of magnitude.\cite{KrauseRehberg1999} In addition, the energy resolution is theoretically improved by a factor of $\sqrt{2}$.\cite{MacDonald1977}\\
Magnesium based alloys, for example the commonly used AZ31, which consists of Mg with 3 wt.\ \% Al and \mbox{1 wt.\ \% Zn}, experience an increasing interest in industrial applications due to their low density and high rigidity. These parameters are not only crucial in car industry and aerospace engineering where energy efficiency enforces the use of light materials but also for the development of lightweight but durable portable electronic devices such as digital cameras, laptops and cell phones. Nevertheless the low plasticity of Mg-based alloys is one major drawback which prevents these materials from being more commonly used.\cite{Skubisz2005}
The macroscopic properties of these materials are strongly correlated with the microscopic structure such as grain size, precipitates and particularly with defects and their chemical vicinity. CDBS is an ideal tool to investigate the latter, because changes of the stoichiometry in the vicinity of open volume defects result in variations of the characteristic electron momentum distribution.\\
However, Mg and its alloys are challenging materials for defect spectroscopy with positrons since it has been shown, that in Mg the trapping rate is relatively low \cite{Hood1982,Seeger1973} and the trapping sites are shallow.\cite{Jackman1974} Whereas the defect induced variation of the Doppler broadening of the positron annihilation line, which is determined by the so called S-parameter, usually lies in the range of $4-5\,\%$ in metals such as Cu or Al, thermally induced defects in Mg only cause $0.6\,\%$.\cite{Segers1980}\\
A standard technique to create defects in metals is the implantation of ions. The near equilibrium solubility of atoms in light metals and the high mobility of point defects \cite{Williams1986} lead to the educated guess, that ion bombardment of AZ31 with Mg-ions could cause the formation of defect clusters with a chemical vicinity deviant from the bulk material. A similar effect has been observed for reactor pressure vessel steels, where the irradiation with fast neutrons induces mobile defects, which encounter Cu atoms and form vacancy-Cu complexes.\cite{Nagai2001}\\
CDBS allows the distinction between elements and therefore the investigation of binary alloys as described above. In contrary to that, the investigation of defect containing ternary alloys such as ion irradiated AZ31 with CDBS is not straight forward, since the elemental contribution of three different elements and the changes in the electron momentum distribution due to defects have to be thoroughly separated. In order to measure changes of the stoichiometry in the vicinity of defects in AZ31, samples of pure Mg have been irradiated with Mg-ions in order to create exclusively defects and to observe the respective change of the $511\,\mathrm{keV}$-annihilation line. Another set of Mg-samples has been irradiated with Al- and Zn-ions to investigate how the shape of the annihilation line changes, if these atoms agglomerate in the vicinity of open volume defects. A comparison with Mg-ion irradiated AZ31-samples should then reveal possible changes of the stoichiometry in the vicinity of defects.\\
This paper is organized as follows. In section \ref{CDBatNEPOMUC}, we describe the CDBS-facility at the intense positron source NEPOMUC. Then the sample preparation is presented in section \ref{SamplePreparation} together with simulations of the resulting vacancy distribution, followed by a description of the required measurement parameters in section \ref{Measurements}. Afterwards the obtained experimental results are presented and discussed, namely standard depth and laterally resolved DB as well as CDB measurements in sections \ref{S(E)DB}, \ref{S(xy)DB} and \ref{CDB}. The paper is concluded in section \ref{Conclusion}.

\section{Experiments}\label{Experiments}

\subsection{The CDB-spectrometer at NEPOMUC}\label{CDBatNEPOMUC}
The energy of the positron beam provided by NEPOMUC \cite{Hugenschmidt2005,Hugenschmidt2006}, the high intense positron source at {FRM II} in Garching (Germany), can be adjusted between 15 and $1000\,\mathrm{eV}$. For measurements with the CDBS, the beam energy is set to $1\,\mathrm{keV}$. A longitudinal magnetic field guides the beam through an evacuated beam tube to the CDB-spectrometer, which has an aperture at the entrance in order to reduce the beam diameter from 20 to 3\,mm. Therefore the beam intensity is decreased from $5 \cdot 10^8$ to roughly $5 \cdot 10^6$ positrons per second, otherwise the energy resolution of the high purity germanium detectors would deteriorate due to too high count rates. In addition the collimation facilitates the subsequent focusing of the beam onto the sample. Afterwards, the beam is magnetically guided to the analysis chamber and non adiabatically released from the longitudinal field by a magnetic field termination. Subsequent electrical lenses focus the divergent positron beam onto the sample.\cite{Stadlbauer2006} A focus with $2.5\,\mathrm{mm}$ in x- and $1.5\,\mathrm{mm}$ in y-direction has been achieved.\cite{Stadlbauer2007}\\
Samples of $20 \times 20\,\mathrm{mm^2}$ can be moved in x- and y-direction perpendicular to the beam axis in order to perform two dimensional scans. The sample potential is adjustable between 0 and $30\,\mathrm{kV}$, thus resulting in a positron energy of 1 to $31\,\mathrm{keV}$, in order to adjust the mean penetration depth of the positrons up to several $\mathrm{\upmu m}$ which is given by the Makhovian profile. Two high purity germanium detectors in collinear geometry with an efficiency of 30\% measure the annihilation radiation emitted from the sample with an energy resolution between 1.2 and $1.4\,\mathrm{keV}$ at $477.6\,\mathrm{keV}$ (this gamma energy is emitted from $\mathrm{^7Be}$ and used as a monitor line). Single photopeak count rates up to $2 \times 3000\,\mathrm{s}^{-1}$ are achievable without major decrease in energy resolution. The signal processing is performed by two digital signal processors from Canberra (DSP 2060) \cite{Koskelo1999} which are connected to the multiparameter system MPA3 from FastComTec.\cite{FastMPA3}

\subsection{Sample Preparation}\label{SamplePreparation}
Samples of pure Mg and AZ31 have been used for the investigations of the present work. From the as received materials, pieces with $20 \times 20 \times 3\,\mathrm{mm^3}$ were produced which fit into the sample holder of the CDB-spectrometer at NEPOMUC.\cite{Stadlbauer2006} Every sample has been polished with 4000 grit abrasive SiC-paper. Subsequently the AZ31-samples were Syton-polished ($\mathrm{SiO_2}$-particles), whereas the Mg-samples were polished with diamond polish paste (grainsize between 1 and $3\,\mathrm{\upmu m}$). This polishing procedure may cause defects at the surface. Nevertheless, the presented measurements are still valid, since the bulk properties of the samples remain unchanged, as we have observed in energy dependent DB-measurements (see section \ref{S(E)DB}). In order to ensure, that all samples are defect free (despite the surface), they have been annealed in a $\mathrm{CO_2}$-atmosphere with a pressure of $1\,\mathrm{bar}$ for $1\,\mathrm{h}$ before the ion irradiation. The temperature was set to $693\,\mathrm{K}$ for the Mg-samples, whereas the AZ31-samples were annealed at $653\,\mathrm{K}$. For temperatures higher than $400\,\mathrm{K}$ even defect clusters are completely removed in Mg.\cite{LandoltBoernstein1991} Rapid quenching of the samples after annealing as a solution treatment has been avoided, since this treatment could lead to thermally induced formation of defects which then may remain present. The grain size of annealed AZ31 varies between $50\,\mathrm{\upmu m}$ and $700\,\mathrm{\upmu m}$ in contrast to rapidly quenched AZ31.\cite{Pravdic2005} The annealing of AZ31 without quenching leads to a distinct grain growth.\cite{PerezPrado2002} Nevertheless, due to this annealing procedure, the positron annihilations will take place in a solid solution of Mg containing Al and Zn at their room temperature equilibrium concentration and do not represent the nominal composition of the alloy.

\begin{figure}[h]
\center
\includegraphics[height=2.2in]{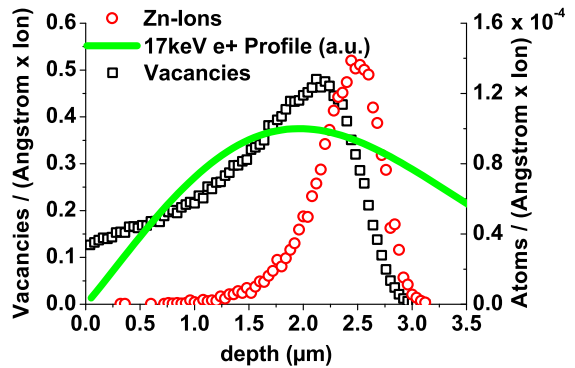}
\caption{Vacancy and Zn-ion distribution in Mg as simulated with SRIM (energy of the ions: $3\,\mathrm{MeV}$). The solid line represents the Makhovian implantation profile of $17\,\mathrm{keV}$-positrons.}
\label{mstadlba_fig1}
\end{figure}

These Zn-containing grains are surrounded by a so called perlit phase, which consists of $\mathrm{Mg}_{17} \mathrm{Al}_{12}$. Since the positron diffusion length is orders of magnitudes smaller than the grain size, the influence of the grain boundaries trapping probability is negligible.\\
The penetration depth of positrons is given by the Makhovian profile for low energies:\cite{PuskaNieminen1994}

\[\bar{z} = \frac{A}{\rho} \cdot E^n_+\]

where $A$ and $n$ are theoretically determined material dependent parameters, and E represents the energy of the positrons in keV. In the present work the values for aluminum $A = 3.7\,\mathrm{\upmu g /cm^2 keV^{-n}}$ and $n = 1.67$ have been used.\cite{PuskaNieminen1994} With the density $\rho = 1.74\,\mathrm{g / cm^3}$ of magnesium the mean penetration depth of positrons lies in the range between 0.02 and $6.3\,\mathrm{\upmu m}$ for positron energies between 1 and $31\,\mathrm{keV}$ covered by the CDB spectrometer at NEPOMUC.\\
The ion irradiation was performed at the $3\,\mathrm{MeV}$-Tandetron in Rossendorf,\cite{Friedrich1996} where a maximum mean implantation depth for Zn-ions into Mg and AZ31 of $2.3\,\mathrm{\upmu m}$ was achievable as shown in figure \ref{mstadlba_fig1}. The depth range of this vacancy and ion distribution is covered with the CDBS, since the sample potential is adjustable between 0 and $-30\,\mathrm{kV}$. The figure was obtained by simulating the ion and vacancy distribution in a Mg-sample with the program SRIM-2003.\cite{Ziegler2004} Additionally the Makhovian profile for positrons with $17\,\mathrm{keV}$ is plotted as a solid line. In order to achieve the same ion distribution for the Mg- and Al- ions, the respective energy of the ion beam was set to 1.4 and $1.6\,\mathrm{MeV}$ (see table \ref{samples}).

\begin{figure}[h]
\center
\includegraphics[height=2.2in]{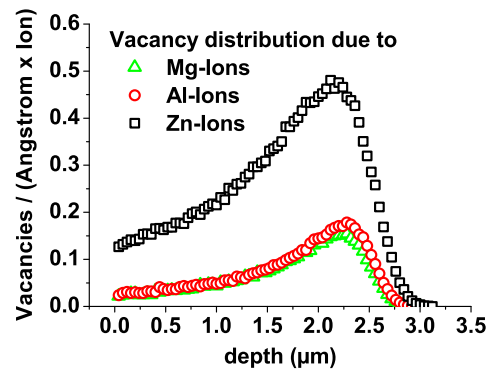}
\caption{Vacancy distribution in Mg as simulated with SRIM after irradiation with Mg-, Al- and Zn-ions. The Zn-ions cause the highest vacancy concentration.} 
\label{mstadlba_fig2}
\end{figure}

\begin{figure}[h]
\center
\includegraphics[height=1.2in]{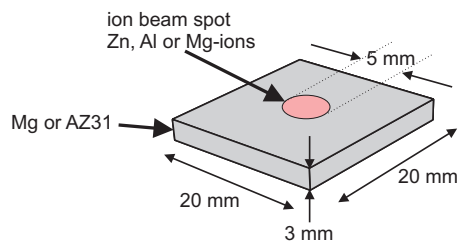}
\caption{Schematic drawing of the ion irradiated samples. The polished and annealed Mg-samples were irradiated with Al-, Zn- and Mg-ions, whereas the AZ31 sample was irradiated only with Mg-ions. The dose was $3 \cdot 10^{14}\,\mathrm{ions/cm^{-2}}$.}
\label{mstadlba_fig3}
\end{figure}

\begin{figure}[h]
\center
\includegraphics[height=2.5in]{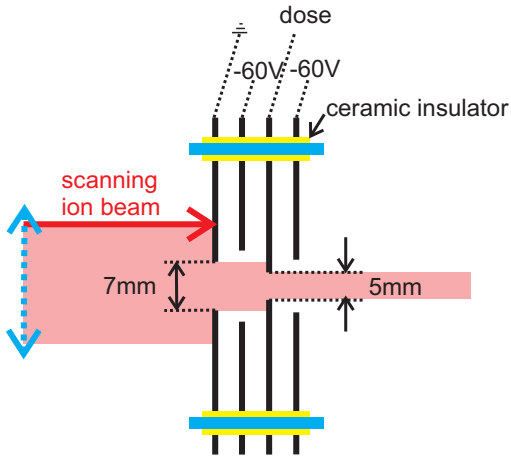}
\caption{Schematic view of the 4-plate aperture. The different sizes of the holes in the first and third plate allow a reliable dose measurement, while the second and the fourth plate reduce the secondary electron background.}
\label{mstadlba_fig4}
\end{figure}

Since the CDBS-facility at NEPOMUC allows spatially resolved measurements with a lateral resolution of approx. $2\,\mathrm{mm}$, the diameter of the ion beam was reduced by an aperture to $5\,\mathrm{mm}$ in order to irradiate exclusively the central region of the samples (figure \ref{mstadlba_fig3}). However, with this small beam diameter the ion dose could not be measured with the permanently installed faraday cup in the implantation chamber of the Tandetron for the described samples, since it has been optimized for scanning the ion beam over wafers with a diameter of $50.8\,\mathrm{mm}$ for implanting and doping purposes. Using a simple aperture with a $5\,\mathrm{mm}$ hole and adjusting the scanning area of the ion beam such that the faraday cup is included would increase the necessary beam time. Therefore an aperture was constructed which consists of 4 plates with 7-, 10-, 5- and $8\,\mathrm{mm}$-holes (see figure \ref{mstadlba_fig4}). The different sizes of the holes in the first and third plate allow a reliable dose measurement via measuring the current at the third plate. The second and the fourth plate are used to screen secondary electrons by applying a potential of $-60\,\mathrm{V}$.

For the present work a set of four samples with $3 \cdot 10^{14}\,\mathrm{ions/cm^{-2}}$ has been produced. In table \ref{samples} these irradiated samples except the reference materials are listed together with labels, which are used in the text.

\begin{table}[h]
\caption{List of all described samples except the reference materials with labels, which are used in the text. The ion dose was $D=3 \cdot 10^{14}\,\mathrm{ions/cm^2}$.}
\label{samples}
\begin{tabular}{ccc} \hline
Ion Type                                   & Sample & Label \\ \hline
$\mathrm{Mg^+}$ ($1.4\,\mathrm{MeV}$)      &  Mg    & Mg/Mg \\
$\mathrm{Al^+}$ ($1.6\,\mathrm{MeV}$)      &  Mg    & Al/Mg \\
$\mathrm{Zn^+}$ ($3.0\,\mathrm{MeV}$)      &  Mg    & Zn/Mg \\
$\mathrm{Mg^+}$ ($1.4\,\mathrm{MeV}$)      &  AZ31  & Mg/AZ31 \\ \hline
\end{tabular}
\end{table}

Similar sample sets were irradiated with other doses between $3 \cdot 10^{13}$ and $3 \cdot 10^{16}\,\mathrm{ions/cm^2}$. The measurements on the samples with the maximum dose are presented elsewhere.\cite{Stadlbauer2007}

\section{Measurements and Discussion}\label{Measurements}
For conventional Doppler broadening measurements (DB) a single high purity germanium detector is used to measure the width of the annihilation line. A standard parameter is the so-called S-parameter which measures the width of the annihilation line by integrating an arbitrarily chosen central region and dividing it by the overall integral of the photopeak. Since the S-parameter is mainly sensitive to the low momentum part of the annihilation line, it is a very sensitive parameter for defect investigations.\cite{Coleman2000}\\
In order to analyze the shape of the annihilation line in more detail, two germanium detectors in coincidence have to be used (CDBS). The Compton background is almost completely suppressed by recording the energy of both emitted $\gamma$-quanta simultaneously and taking only events into account with a sum energy of $1022\,\mathrm{keV}$.\cite{KrauseRehberg1999} The different shapes of the spectra appear very clearly, if the $511\,\mathrm{keV}$-peaks are normalized to equal area and divided by a reference spectrum. These curves are denoted as ratio curves or signatures.\\
Firstly, DB-measurements have been performed of all samples, as a function of the positron energy, i.e. the S-parameter S(E) reveals depth dependent information within the Makhov-profile. Secondly, spatially resolved two dimensional S-parameter scans with respect to the position of the positron beam on the sample S(x,y) with a stepsize of $1\,\mathrm{mm}$ have been accomplished. After that, each sample has been investigated with CDBS by using the two opposite germanium detectors in coincidence and recording the shape of the annihilation line with suppressed background. The positron energy for the latter measurements was set to $17\,\mathrm{keV}$, which matches best the applied implantation depth of the ions.

\subsection{DB-Measurements as a Function of the Positron Energy - S(E)}\label{S(E)DB}
\begin{figure}[ht]
\center
\includegraphics[height=2.2in]{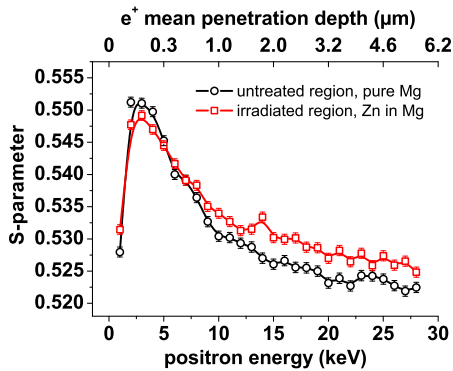}
\caption{S-parameter as a function of the positron energy measured for the Zn/Mg-sample in comparison to untreated pure Mg. The two curves differ at the surface (between 1 and roughly $4\,\mathrm{keV}$) and for greater positron penetration depths (between 9 and $28\,\mathrm{keV}$). This behavior is attributed to the irradiation induced defect distribution as shown in figure \ref{mstadlba_fig1}. The solid lines serve as an eye guide.}
\label{mstadlba_fig5}
\end{figure}

Measurements of the S-parameter as a function of the positron energy in the irradiated area as well as in the untreated region were performed for all samples, listed in table \ref{samples}. The energy resolution at $477.6\,\mathrm{keV}$ was $1.3\,\mathrm{keV}$. For the $511\,\mathrm{keV}$ annihilation line the photopeak count rate was typically $2500\,\mathrm{cts/s}$ (total count rate: $14000\,\mathrm{cts/s}$). Hence $5 \cdot 10^5\,\mathrm{counts}$ were collected during a measurement time of $3\,\mathrm{min}$. Figure \ref{mstadlba_fig5} compares the untreated with the Zn-ion irradiated region of the Zn/Mg-sample. The low S-parameter at $1\,\mathrm{keV}$ positron energy was found to be dependent on the polishing procedure and not further quantified, whereas the high S-parameters for energies between 2 and $10\,\mathrm{keV}$ are a consequence of back diffusing positrons which form positronium.\cite{Triftshaeuser1982} Positronium formation is possible in non-metals \cite{Ferrel1956} or at surfaces.\cite{Ishii1987}

The S-parameter differs at the sample surface (between 1 and roughly $4\,\mathrm{keV}$) and for greater positron penetration depths (between 9 and $29\,\mathrm{keV}$). This is ascribed to defects which have been created during the irradiation procedure. Since the defect profile stretches to the surface as shown in figure \ref{mstadlba_fig1}, the positrons are trapped in defects and hence detained from diffusing back to the surface which reduces the rising of the S-parameter at low sample potentials, i.e. low energies and therefore very small positron penetration depths. On the other hand, the defects in the bulk region result in a higher S-parameter compared to the untreated bulk region, since defects cause a smaller line shape and hence an increase of the S-parameter.\cite{Coleman2000}\\
The difference between the two curves reaches its maximum between 13 and $21\,\mathrm{keV}$, whereas for higher positron energies it starts to decrease, which is ascribed to the shape of the vacancy distribution as shown in figure \ref{mstadlba_fig1} with a maximum defect concentration  at $2.3\,\mathrm{\upmu m}$ equal to $17\,\mathrm{keV}$ positron energy. The other irradiated samples show a similar behavior even though with a smaller difference between the two curves.

\subsection{Laterally Resolved DB-Measurements - S(x,y)}\label{S(xy)DB}
\begin{figure*}[ht]
\center
\includegraphics[height=2.2in]{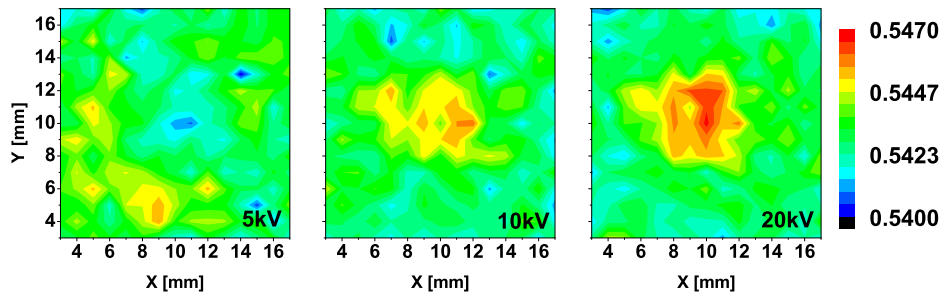}
\caption{S-parameter as a function of the position of the positron beam on the Zn/Mg-sample at various sample potentials (step size $1\,\mathrm{mm}$). The minimum S-parameter for every picture was set to 0.54 in order to use an identical color scale which ranges from 0.54 to 0.547 in arbitrary units. At $5\,\mathrm{kV}$ the ion beam spot is not detectable, whereas at $10\,\mathrm{kV}$ the irradiated region already becomes visible (max. S-deviation $1\,\%$). For $20\,\mathrm{kV}$ the ion beam spot is clearly reproducible as expected from the above described S(E)-measurements.} 
\label{mstadlba_fig6}
\end{figure*}

In order to image the irradiated region on the sample, the S-parameter has been recorded depending on the position of the positron beam for all samples at various potentials. A region of $15 \times 15\,\mathrm{mm^2}$ has been scanned with a step size of $1\,\mathrm{mm}$. The measurement time per point was $3\,\mathrm{min}$ with a count rate of $2800\,\mathrm{cts/s}$ in the $511\,\mathrm{keV}$-photopeak resulting in the same energy resolution as in the above mentioned S(E)-measurements. The result for the Zn/Mg-sample is shown in figure \ref{mstadlba_fig6}, for which the difference between the irradiated and the untreated region is very distinct. The minimum S-parameter value is set to 0.54 for every picture and the color map ranges up to 0.547. At $5\,\mathrm{kV}$ sample potential ($6\,\mathrm{keV}$ positron energy) the irradiated region is almost not visible besides an area with slightly decreased S-parameter as expected from the S(E)-measurements (figure \ref{mstadlba_fig5}). For $10\,\mathrm{kV}$ sample potential the area affected by the ion beam spot is clearly detectable. The maximum S-parameter deviation in this picture amounts to $1\,\%$, whereas for $20\,\mathrm{kV}$ the irradiated region is even more distinctly reproducible with a maximum S-parameter contrast of $1.3\,\%$. Also the diameter of the irradiated area matches the size of the ion beam of $5\,\mathrm{mm}$ within the spatial resolution of the CDB-spectrometer. Although the trapping rate of positrons in Mg is rather small and the trapping sites are shallow,\cite{Seeger1973,Jackman1974,Hood1982} the ion beam spot on the samples has been clearly reproduced.

\subsection{Coincident Doppler Broadening Measurements}\label{CDB}
\begin{figure}[ht]
\center
\includegraphics[height=2.2in]{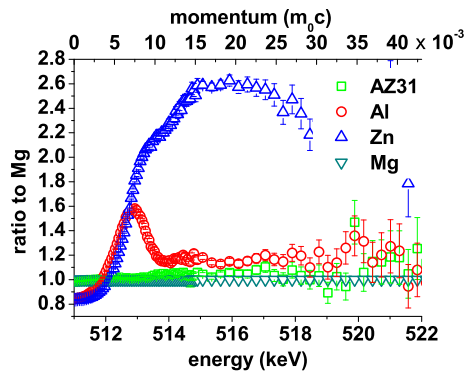}
\caption{CDB ratio curves of pure and annealed Al, Zn and AZ31 relative to Mg at a sample potential of $25\,\mathrm{kV}$ for Mg, Al, AZ31 and $28\,\mathrm{kV}$ for Zn. The elements are clearly distinguishable, since Al and Zn show increased intensities in the high momentum region ($> 10 \cdot 10^{-3}\,\mathrm{m_o c})$. The Zn-signature dominates this region.}
\label{mstadlba_fig7}
\end{figure}

\begin{figure}[ht]
\center
\includegraphics[height=2.2in]{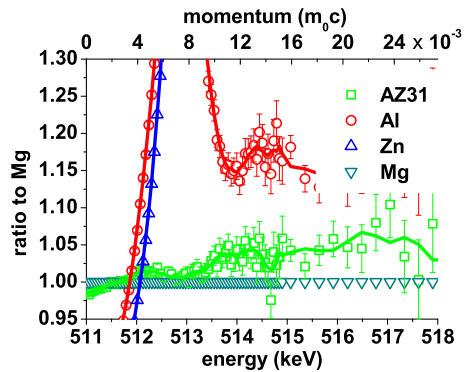}
\caption{Zoom into figure \ref{mstadlba_fig7}. The AZ31-signature reveals a higher intensity for momenta greater than $5 \cdot 10^{-3}\,\mathrm{m_0c}$ which is attributed to the contribution of Zn- and Al-atoms. The solid lines serve as an eye guide.}
\label{mstadlba_fig8}
\end{figure}

Firstly, polished and annealed samples of pure Mg, Al, Zn and AZ31 have been investigated with CDBS in order to record the characteristic shape of the annihilation line and in particular the high momentum region for each material. A sample potential of $25\,\mathrm{kV}$ for the light materials ensured, that all positrons annihilate in the bulk, whereas the sample potential for the Zn-measurement was set to $28\,\mathrm{kV}$, because S(E)-measurements on annealed Zn showed, that at $25\,\mathrm{kV}$ the bulk S-parameter was not fully reached. In order to increase statistics, both sides of the symmetric photopeak are used for data treatment, i.e. the red shifted side ($E < E_0=511\,\mathrm{keV})$ of the peak is mirrored at $E_0$ and added to the blue shifted side ($E > E_0$). Afterwards these curves are normalized to equal integrated intensity of the photopeak and divided by the reference spectrum (for example annealed Mg). The error bars are determined by applying the Gaussian error propagation law to the described data treatment. In order to minimize the large error bars due to low statistics for momenta higher than $15 \cdot 10^{-3}\,\mathrm{m_0c}$, the bin size is increased from $56.7\,\mathrm{eV/bin}$ to $283.5\,\mathrm{eV/bin}$. The spectra contain $1 - 2 \cdot 10^7\,\mathrm{counts}$ and were recorded with $1.3\,\mathrm{keV}$ energy resolution at $477.6\,\mathrm{keV}$ for each detector.\\
In figure \ref{mstadlba_fig7} the CDB-ratio curves (signatures) of the annealed pure elements Al, Zn and the annealed alloy AZ31 are shown with respect to annealed Mg as a reference. The signatures of Zn and Al are clearly distinguishable from Mg, especially Zn reveals a large intensity for momenta higher than $7 \cdot 10^{-3}\,\mathrm{m_0c}$. The distinct peak in the Al-curve at $6.6 \cdot 10^{-3}\,\mathrm{m_0c}$ or $512.7\,\mathrm{keV}$ is a consequence of the mismatching Fermi energies of Al and Mg ($7.08$ and $11.7\,\mathrm{eV}$ respectively) which lead to a Doppler shift of $1.3$ and $1.7\,\mathrm{keV}$. A zoomed view of figure \ref{mstadlba_fig7} is plotted in figure \ref{mstadlba_fig8}, which shows a slight deviation of the AZ31-signature with respect to pure Mg towards Zn and Al. This deviation is attributed to the presence of Zn and Al-atoms in the annealed AZ31. 

\begin{figure}[ht]
\center
\includegraphics[height=2.2in]{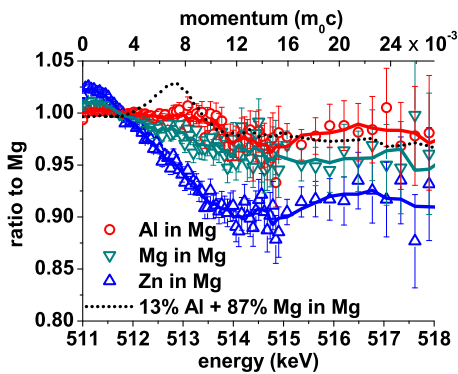}
\caption{CDB ratio curves of the Mg-, Al- and Zn-ion irradiated samples with untreated Mg as a reference at $25\,\mathrm{kV}$ sample potential. The Zn in Mg signature is more pronounced which is attributed to a higher concentration of vacancy clusters. The dotted line represents a linear combination with $13\%$ of the pure Al-spectrum and $87\%$ of the Mg/Mg-spectrum (see text). The solid lines serve as an eye guide.} 
\label{mstadlba_fig9}
\end{figure}

Figure \ref{mstadlba_fig9} presents the three CDB-curves of the irradiated Mg/Mg-, Al/Mg- and Zn/Mg-samples relative to untreated Mg. All three signatures reveal a reduced contribution of high momentum electrons, as expected due to the irradiation induced defects \cite{Coleman2000}. Hence a decline between 5 and $18 \cdot 10^{-3}\,\mathrm{m_0c}$ is observed with respect to the reference spectrum of annealed Mg. The signatures of the Mg/Mg- and the Al/Mg-sample are similar and the Zn/Mg-sample reveals the most distinct deviation. This could be ascribed to the irradiation induced vacancy distribution, which was calculated with SRIM (see figure \ref{mstadlba_fig2}) in advance. The differences from unity of the Mg/Mg and the Zn/Mg-curve scale approximately by a factor of two which is at least roughly represented by the simulated vacancy distributions. Nevertheless, SRIM only computes the initial vacancy distribution without taking annealing effects into account. Since monovacancies in Mg anneal already below room temperature, but vacancy clusters survive up to $400\,\mathrm{K}$, deviations occur from the direct proportionality of the simulated vacancy distributions and the differences of the measured ratio curves.\cite{LandoltBoernstein1991} For a quantitative interpretation, positron lifetime measurements have to be performed in future to reveal the concentration and type of vacancies.

\begin{figure}[ht]
\center
\includegraphics[height=2.2in]{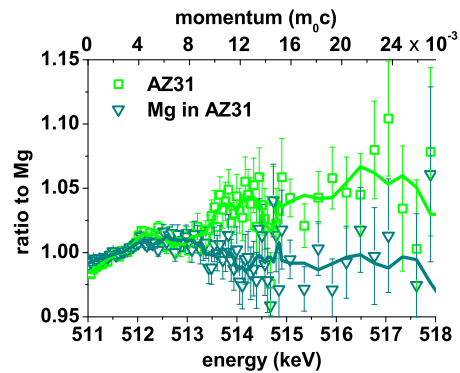}
\caption{CDB ratio curve of annealed AZ31 and the Mg/AZ31-sample normalized to annealed Mg. The irradiated sample reveals a defect induced decline of the curve in the high momentum region as expected. The solid lines serve as an eye guide.}
\label{mstadlba_fig10}
\end{figure}

\begin{figure}[ht]
\center
\includegraphics[height=2.2in]{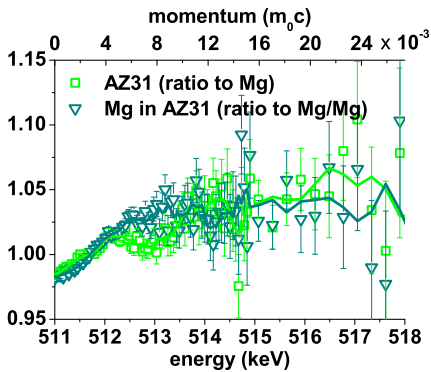}
\caption{CDB ratio curve of annealed AZ31 (with annealed Mg as a reference) and the Mg/AZ31-sample (with Mg/Mg as a reference). Both signatures are identical within the statistical error. For interpretation see text. The solid lines serve as an eye guide.}
\label{mstadlba_fig11}
\end{figure}

The Makhov-Profile as plotted in figure \ref{mstadlba_fig1} shows, that $70\,\%$ of the positrons annihilate outside the ion distribution and hence only $30\,\%$ of the implanted positrons annihilate in a region, where the ion-distribution is not zero. Nevertheless, if the implanted ions accumulated in the vicinity of the created defects, the trapped positrons should also annihilate with the respective core electrons and therefore the characteristic signatures should be detectable with CDBS. Due to the similar masses of Al- and Mg-ions, the resulting vacancy distributions are nearly the same according to the SRIM-simulations and hence, a similar behavior of the Al/Mg and the Mg/Mg-curve in the high momentum region is expected. Nevertheless, due to a significant difference in this region, a decoration of the irradiation induced defects with Al-atoms can not be excluded. In order to investigate this observation in more detail, the Mg/Mg-spectrum has been linearly combined with $13\%$ of the annealed Al-spectrum such that the resulting curve matches the high momentum region of the Al/Mg-curve (dotted line in fig. \ref{mstadlba_fig9}). The pronounced peak between $4$ and $10 \cdot 10^{-3}\,\mathrm{m_0c}$ in the linearly combined curve results not from confinement of the positrons in defects, but from the mismatch of the Fermi energies as explained above, since no significant confinement peaks are visible in none of the three curves.\cite{Calloni2005} Due to the appearance of this peak, a quantitative analysis would not lead to reliable results.
However, the signature of Zn/Mg does not reveal any contribution of localized Zn-electrons, since no increase for high electron momenta is observed. Moreover, its shape is similar to the Mg/Mg-curve within the statistical accuracy and only the amplitude of the deviation varies. This remarkable fact leads to the conclusion, that no Zn-ions are agglomerating in the vicinity of the irradiation induced defects. Consequently, a defect-independent distribution of the implanted ions is assumed since CDBS does not detect the respective signature due to the rather low concentration of the implanted ions. The ion concentration in the respective region amounts to $2 \cdot 10^{18}\,\mathrm{cm^{-3}}$, which equals $0.005\,\mathrm{at}\,\%$ only and is hence not measurable with CDBS, since in annealed AZ31 the room temperature equilibrium concentrations are barely detectable.\\
Figure \ref{mstadlba_fig10} shows the CDBS-signatures of the Mg/AZ31-sample and annealed AZ31 with annealed Mg as a reference. Both curves are identical within the error bars from 2 up to $8 \cdot 10^{-3}\,\mathrm{m_0c}$. For higher momenta the irradiated sample reveals a distinct decline due to the created defects.\\
In order to account for the contribution of defects to the annihilation line of the Mg/AZ31-sample, the spectrum of Mg/AZ31 has been normalized to the Mg/Mg-spectrum. This normalization cancels out the contribution of defects in pure Mg in the high momentum region. The respective curve is plotted in figure \ref{mstadlba_fig11} together with the signature of annealed AZ31 with annealed Mg as a reference. It is clearly visible, that for momenta higher than $10 \cdot 10^{-3}\,\mathrm{m_0c}$ the two curves are the same within the statistical error. The visible effects for smaller momenta are resulting from changes in the momentum distribution of the conducting electrons due to varying defect concentrations and can therefore not be used to characterize the chemical vicinity of defects as explained above.

Thus, the high momentum part of the ratio curve between annealed AZ31 and Mg does not change significantly, if both samples are irradiated with the same dose of Mg-ions. This behavior shows, that neither Zn- nor Al-atoms agglomerate in the vicinity of ion irradiation induced defects in AZ31, otherwise a significant increase in the high momentum region would be detectable, since the high momentum parts of the Zn- and Al-curves reveal higher intensities than Mg as shown in figure \ref{mstadlba_fig7}. Consequently, within the accuracy of the measurement, the chemical vicinity of crystal defects produced by Mg-ion irradiation remains unchanged compared to the annealed bulk material. Despite the high defect mobility in light metals, no ion-irradiation induced change of the stoichiometry in the vicinity of defects was found.

\section{Conclusion and Outlook}\label{Conclusion}
This work presented Doppler broadening and coincident Doppler broadening measurements of Mg- and AZ31 samples irradiated with Mg-, Al- and Zn-ions. The irradiated region in the center of the samples has been investigated with laterally resolved Doppler broadening (DB) measurements in order to image the ion beam spot. Although the trapping rate of positrons in Mg is rather small and the trapping sites are shallow,\cite{Seeger1973,Jackman1974,Hood1982} the ion beam spot on the samples has been clearly reproduced within the spatial resolution of the CDBS-facility. As expected from simulations with the program SRIM, which was used to calculate the ion and vacancy distribution, the Doppler broadening of the annihilation line depends significantly on the positron penetration depth, which was adjusted up to $5.5\,\mathrm{\upmu m}$.\\
Coincident Doppler broadening measurements on the pure and annealed metals Mg, Al and Zn revealed clearly separable CDB-ratio curves. Moreover the ratio curve of AZ31 is distinguishable from pure and annealed Mg due to the small fractions of Al and Zn.\\
Ion irradiated Mg has been examined with CDBS in order to investigate, whether the irradiation induced defects in annealed Mg form vacancy clusters and furthermore, whether the implanted ions are agglomerating in their vicinity. It was shown, that the implantation with Mg- and Al-ions lead to a less distinct change of the shape of the annihilation line compared to the irradiation with Zn-ions. This was explained qualitatively with the aid of SRIM-simulations, which clearly showed, that the Zn-ions create a two times higher integrated vacancy concentration due to their larger mass.
The characteristic signature of Zn was not observed in these spectra which leads to the conclusion, that within the accuracy of the measurement the implanted Zn-ions do not agglomerate in the vicinity of the defect clusters, whereas a decoration of the irradiation induced defects with Al-atoms can not be excluded.\\
The Mg-ion irradiation induced defects in AZ31 were analyzed with CDBS as well. The ratio curves distinctively revealed, that the defects lead to a change in the shape of the annihilation line. In order to account for this defect contribution, the annihilation line of the Mg-ion irradiated AZ31 was normalized to Mg-ion irradiated Mg, which exclusively contains defects with no other elemental signature. The resulting ratio curve is in the high momentum region (above the Fermi energy where only core electrons contribute to the annihilation process) no longer distinguishable from the ratio curve of annealed AZ31 normalized to annealed Mg. In contrary to reactor pressure vessel steels, where neutron damage leads to Cu aggregations in the Fe matrix,\cite{Nagai2001} this fact shows, that in annealed AZ31 neither Al nor Zn agglomerate in the vicinity of the ion irradiation induced defects within the accuracy of the measurement, despite the high mobility of defects in light metals.\\
Since it is not straight forward to distinguish between different defect types and the absolute defect concentration with Doppler broadening measurements, positron lifetime spectroscopy should be applied. The pulsed low energy positron system PLEPS \cite{BauerKugelmann2001} is being currently installed at the positron source NEPOMUC and will be used for further examinations on the samples described in this paper. Furthermore a cryostat is presently installed in the sample chamber of the CDBS-facility at NEPOMUC, in order to measure shallow positron traps with increased sensitivity at low temperatures.\\
Ab-initio computations of CDB-spectra have already been successfully performed for example for vacancies and vacancy-oxygen complexes in silicon,\cite{Kuriplach1998} Cu precipitates in Fe,\cite{Nagai2004} vacancies and vacancylike defects in Al \cite{Calloni2005} and for vacancies and vacancy-solute complexes in Al-alloys.\cite{Folegati2007} Due to the hcp-structure of Mg and the presence of vacancy clusters in the samples described in the present work, such computations are more complex.\cite{Folegati2007_a} Nevertheless, the calculation of the electron momentum distribution in Mg and AZ31 with point defects and vacancy clusters may yield a more theoretical understanding of the observed spectra.

\begin{acknowledgments}
The authors would like to thank Dr. Friedrich and his team from the FZR Rossendorf in Dresden for making the 3MeV-Tandetron available and for their very helpful support during the beam time. Furthermore the authors gratefully acknowledge Dr. Matz Haaks from HISKP in Bonn for programming and providing the S-parameter analysis tool MSPEC and MSPEC2D for obtaining the projection from the  two dimensional CDB-spectra.
\end{acknowledgments}

\bibliography{IrrCDB}

\end{document}